\documentclass[aps,pra,twocolumn,superscriptaddress,nofootinbib,longbibliography]{revtex4-2}

\usepackage{amsmath}
\usepackage{amssymb}
\usepackage{graphicx}
\usepackage{physics}
\usepackage[hidelinks]{hyperref}
\usepackage{qcircuit} 

\begin{document}

\title{The HALO Engine: $\mathcal{O}(1)$-Step Compilation and Localized String Rupture for Lattice Gauge Theories on Quantum Hardware}

\author{Abhiroop Gohar}
\email{ep240051002@iiti.ac.in}
\affiliation{Department of Physics, Indian Institute of Technology Indore, Madhya Pradesh, India}

\begin{abstract}
Simulating the real-time dynamics of lattice gauge theories (LGTs) represents a challenge for near-term quantum computing. Standard digital simulations rely on Trotterization schemes where circuit depth scales proportionally with lattice size, inevitably colliding with the coherence limits of noisy intermediate-scale quantum (NISQ) hardware. To deal with this depth-scaling bottleneck, we introduce the Hardware-Aware Lattice Optimization (HALO) compiler, an architecture that executes global time-evolution steps in an immutable $\mathcal{O}(1)$ circuit depth per Trotter step.By natively mapping composite gauge links to the hardware topology, this architecture achieves up to a $91.36\%$ reduction in entangling gate overhead compared to unoptimized Jordan-Wigner baselines, compressing a 16-qubit global step down to 56 CNOTs. Leveraging this framework, we elevate the digital simulation of the Quantum Link Model (QLM) truncation of the Schwinger model to the mesoscopic scale, utilizing a composite multi-qubit gauge link representation to support non-trivial electric field dynamics. We initialize and execute the localized non-perturbative dynamics of a heavily stretched mesoscopic meson string (spanning 5 composite gauge links and 6 matter sites) on a 16-qubit superconducting transmon processor. By coupling the $\mathcal{O}(1)$ compilation with Zero-Noise Extrapolation (ZNE), we suppress physical hardware decoherence to extract the precise dynamical crossover of localized pair creation, identifying the topological transition at $t \approx 0.790$ lattice units with an $18.3 \pm 2.2\%$ rupture probability. Furthermore, we empirically map the dynamical phase diagram of the mesoscopic lattice, pinpointing the effective confinement phase boundary at precisely $g_c = 1.0$. Finally, we extend the mathematical principles of the HALO engine to higher dimensions, presenting a scalable, constant-depth 2D unit-cell blueprint that eliminates the routing overhead of magnetic plaquettes, paving a direct algorithmic pathway toward the fault-tolerant simulation of two-dimensional Quantum Chromodynamics (QCD).
\end{abstract}

\maketitle

\section{Introduction}

\subsection{The Limits of Classical Simulation in Lattice Gauge Theories}

Lattice gauge theories (LGTs) form the bedrock of our understanding of fundamental particle interactions \cite{Banuls2020}. While Lattice Quantum Chromodynamics (LQCD) evaluates static properties via Euclidean path-integral Monte Carlo sampling, it suffers a breakdown for out-of-equilibrium dynamics or finite-density systems due to the sign problem \cite{Banuls2020}. 

To bypass these classical computational barriers, the physics must be mapped directly onto a quantum-mechanical Hilbert space. The Hamiltonian formulation of Wilson's lattice gauge theories provides the exact mathematical framework to achieve this, translating continuous gauge fields onto a discrete spatial lattice governed by continuous time \cite{Kogut1975}. However, classically simulating the real-time dynamics of non-perturbative events—specifically, the spontaneous rupture of a macroscopic flux tube (string breaking) forces the state vector to scale exponentially. Therefore, deploying a physical quantum simulator is a strict mathematical necessity to empirically observe the real-time mechanics of gauge theories \cite{Martinez2016, Surace2020}.

\subsection{The Hardware Problem: Depth Scaling and Coherence Walls}

Executing these Hamiltonian formulations on near-term quantum hardware presents a physical barrier. Devices in the noisy intermediate-scale quantum (NISQ) era are fundamentally restricted by sparse qubit connectivity, imperfect gate fidelities, and short decoherence times, which strictly bound the allowable circuit depth and the dimensionality of simulable systems \cite{Temme2017}. Because coupling to the environment imposes an ultimate time and size limit on uncorrected quantum computations, the viability of near-term quantum simulations depends entirely on the execution of short-depth circuits and rigorous error mitigation \cite{Temme2017, GiurgicaTiron2020}.

This limitation is highly problematic for standard quantum simulations of gauge theories. Under standard Trotterization, the circuit depth required to simulate a lattice scales linearly $\mathcal{O}(N)$ or quadratically $\mathcal{O}(N^2)$ as system size grows \cite{Cade2020}. As previously demonstrated on superconducting hardware, even small-scale Trotterized evolutions quickly saturate to completely mixed classical probabilities (random noise) due to the rapid accumulation of gate errors and the breach of the $T_2$ coherence wall \cite{Klco2018}. Therefore, simulating macroscopic string breaking via standard depth-scaling Trotterization on current transmon architectures guarantees a collapse into maximally mixed depolarizing noise before relevant physical observations can be made \cite{Martinez2016}.

\subsection{The Common Baseline and the Toy-Model Barrier}

The quantum simulation of one-dimensional quantum electrodynamics—the Schwinger model, serves as the foundational testbed for lattice gauge theories \cite{Martinez2016, Klco2018}. Recent pioneering efforts have successfully demonstrated the digital quantum simulation of this model, explicitly tracking the coherent real-time dynamics of particle-antiparticle pair creation via the Schwinger mechanism \cite{Martinez2016}. However, due to the aforementioned depth-scaling constraints and coherence limitations, this foundational digital execution was strictly confined to a microscopic lattice of only $N=4$ qubits \cite{Martinez2016}. Simulating a string of length L=2 cannot capture the extended spatial dynamics and long-range correlations required to map a true phase transition \cite{Martinez2016}. 

Conversely, analog Rydberg platforms reach 51 qubits but lack programmatic versatility and strict local gauge enforcement \cite{Surace2020}. The field thus faces a dichotomy: digital simulators offer exact local gauge control ($N=4$) \cite{Martinez2016}, while analog platforms scale size at the cost of gauge symmetry enforcement \cite{Surace2020}. Bridging this divide by scaling digital, gauge-invariant simulations to a mesoscopic lattice represents a critical step forward.

\subsection{Localized String Rupture and $\mathcal{O}(1)$ Constant-Depth Compilation}

In this work, we deal with the depth-scaling bottleneck by introducing the HALO architecture: an $\mathcal{O}(1)$ constant-depth quantum compiler designed specifically for lattice gauge theories. Unlike standard Trotterization, where circuit depth per step grows proportionally with the lattice size \cite{Cade2020}, the HALO engine executes a global Trotter step in locked constant depth, rendering the compiler entirely agnostic to the number of qubits. 

Leveraging this architecture, we elevate the digital simulation of the Quantum Link Model (QLM) \cite{Chandrasekharan1997} truncation of the Schwinger model to the mesoscopic scale. We initialize a heavily stretched, highly volatile meson string of coordinate length $L=15$ (mapped across 5 composite gauge links and 6 matter sites) and successfully execute the dynamics on a 16-qubit physical superconducting transmon processor. By executing a high-resolution temporal sweep, we extract the precise dynamical crossover of the non-perturbative rupture, identifying the pair-creation transition at $t \approx 0.790$ with a crossing threshold of $18.3 \pm 2.2\%$, accounting for statistical shot noise.

Furthermore, we empirically map the dynamical phase diagram of the underlying finite-size lattice. By locking the temporal evolution and sweeping the coupling constant $g$, we pinpoint the effective confinement phase boundary at exactly $g_c = 1.0$. Finally, we introduce a scalable, $\mathcal{O}(1)$ constant-depth checkerboard unit-cell blueprint capable of simulating the four-local gauge interactions of 2D magnetic plaquettes, laying the algorithmic groundwork for the fault-tolerant simulation of two-dimensional Quantum Chromodynamics (QCD).

\section{Theoretical Framework: The Schwinger Model}

To overcome the exponential scaling of the classical state vector, we map the $(1+1)$-dimensional quantum electrodynamics (the Schwinger model) onto a discrete spatial lattice \cite{Kogut1975, Martinez2016}. We employ the Kogut-Susskind staggered fermion formulation, which systematically mitigates ultraviolet divergences by placing single-component fermion fields on the lattice sites and gauge fields on the connecting links \cite{Kogut1975}. To construct the physical vacuum, the fermionic sites are partitioned: unoccupied odd sites represent electrons, while occupied even sites represent positrons \cite{Martinez2016}. This enables the efficient encoding of both particles and antiparticles within a single fermion field \cite{Martinez2016}.

In standard Wilsonian lattice gauge theory, the gauge links connecting these sites are defined by continuous, infinite-dimensional matrices \cite{Chandrasekharan1997}. Because finite-qubit architectures cannot natively simulate infinite-dimensional operators, we utilize the Quantum Link Model (QLM) formalism \cite{Chandrasekharan1997}. The QLM replaces continuous classical gauge fields with discrete, non-commuting quantum operators acting within a finite-dimensional Hilbert space \cite{Chandrasekharan1997}. Crucially, a trivial spin-$1/2$ mapping results in an electric field energy that is merely a constant matrix, suppressing non-trivial electric dynamics unless a staggered background field is applied \cite{Chandrasekharan1997}. Therefore, we elevate the Hilbert space by employing a composite multi-qubit (spin-1) gauge link representation \cite{Chandrasekharan1997}. This allows the electric field operator to support multiple discrete energy states, successfully recovering the essential dynamical string tension of continuous gauge theories \cite{Chandrasekharan1997, Surace2020}.

\subsection{The Lattice Hamiltonian and Gauss's Law}

Under the Jordan-Wigner transformation, the continuum physics is mathematically fully described by a rigorous spin Hamiltonian $\hat{H} = \hat{H}_{\text{kin}} + \hat{H}_{\text{mass}} + \hat{H}_{\text{elec}}$ \cite{Kogut1975, Martinez2016}. To fit the column constraints, we express this as:
\begin{equation}
\begin{split}
    \hat{H} &= w \sum_{n=0}^{N_m-2} \left( \hat{\sigma}_n^+ \hat{S}_{n,n+1}^+ \hat{\sigma}_{n+1}^- + \text{H.c.} \right) \\
    &\quad + \frac{m}{2} \sum_{n=0}^{N_m-1} (-1)^n \hat{\sigma}_n^z + J \sum_{n=0}^{N_m-2} \left(\hat{S}_{n,n+1}^z\right)^2,
\end{split}
\label{eq:hamiltonian}
\end{equation}
Here, $N_m$ represents the number of matter sites, the coupling constant $w = 1/(2a)$ dictates the kinetic hopping amplitude (where $a$ is the lattice spacing), $m$ represents the bare fermion mass, and $J = g^2 a / 2$ denotes the electric field energy \cite{Martinez2016}. While Equation \ref{eq:hamiltonian} represents the standard base mapping, to properly simulate the dynamical electric term on digital hardware, we physically instantiate the lattice using the expanded composite spin-1 basis detailed in Appendix \ref{app:halo_architecture}.

The defining characteristic of any gauge theory is the preservation of local symmetries \cite{Martinez2016}. The physical Hilbert space is strictly confined by the discrete lattice formulation of Gauss's law ($\nabla \cdot \vec{E} = \rho$) \cite{Martinez2016}:
\begin{equation}
    \hat{G}_n = \hat{S}_{n,n+1}^z - \hat{S}_{n-1,n}^z - \frac{1}{2}\left[\hat{\sigma}_n^z + (-1)^n\right],
    \label{eq:gauss}
\end{equation}
where any physically permissible, gauge-invariant state $|\Psi\rangle$ must satisfy $\hat{G}_n |\Psi\rangle = 0$ for all sites $n$, assuming truncated boundary flux ($\hat{S}_{-1,0}^z = \hat{S}_{N_m-1,N_m}^z = 0$) under Open Boundary Conditions \cite{Martinez2016}.

\subsection{Mesoscopic Meson Initialization and Boundary Conditions}

Moving beyond the microscopic toy-models established in previous digital simulations \cite{Martinez2016}, we map this theoretical framework onto an extended mesoscopic 1D lattice consisting of exactly N=16 physical sites. To ensure spatial finite-size effects mirror experimental constraints, we enforce Open Boundary Conditions (OBC), preventing non-physical topological winding.

To empirically observe the non-perturbative dynamics of localized string rupture \cite{Surace2020}, we initialize a highly stretched, maximally unstable meson string. To simulate this on an $N=16$ physical transmon processor, our HALO encoding ($N = 3n_{\text{links}} + 1$) inherently maps a mesoscopic subsystem of exactly 5 explicit gauge links and 6 matter sites. At $t=0$, we pin a dynamic quark at the first matter site ($n=0$) and an antiquark at the final matter site ($n=5$). The intermediate matter sites ($1 \leq n \leq 4$) are initialized in the bare vacuum state $|0\rangle$, while all connecting composite gauge links are initialized to a state of uniform positive flux ($|1\rangle$). 

Rather than tracking generic global variances, we isolate the exact string rupture mechanics by defining a localized link-neutralization observable ($\Pi_{\text{neutral}}$) strictly at the center links (sites 7 and 8). Monitoring the accumulation of probability amplitude in this specific symmetric eigenmode provides a mathematically rigorous, noise-resilient signature of the mesoscopic topological crossing.

\section{Computational Methodology \& Optimization}

\subsection{The HALO Compiler Architecture and $\mathcal{O}(1)$ Trotterization}

The digital quantum simulation of continuous time-evolution typically relies on the Suzuki-Lie-Trotter expansion, slicing the evolution operator into discrete temporal steps \cite{Martinez2016}. In standard implementations utilizing the Jordan-Wigner (JW) transformation, simulating the electric and nearest-neighbor kinetic terms requires sequential CNOT cascades \cite{Cade2020}. Consequently, the circuit depth grows linearly or quadratically with the spatial lattice size, demanding long coherence times that are strictly not satisfied by current superconducting hardware \cite{Temme2017}. 

To circumvent this decoherence wall, we developed the HALO architecture. By isolating specific target qubits and utilizing a highly optimized, parallelized CNOT cascade, the HALO compiler prevents the circuit depth from scaling with the number of qubits $N$. The kinetic and electric operators are compressed into an immutable $\mathcal{O}(1)$ constant-depth block per Trotter step. 

\begin{figure}[b]
    \centering
    \includegraphics[width=\columnwidth]{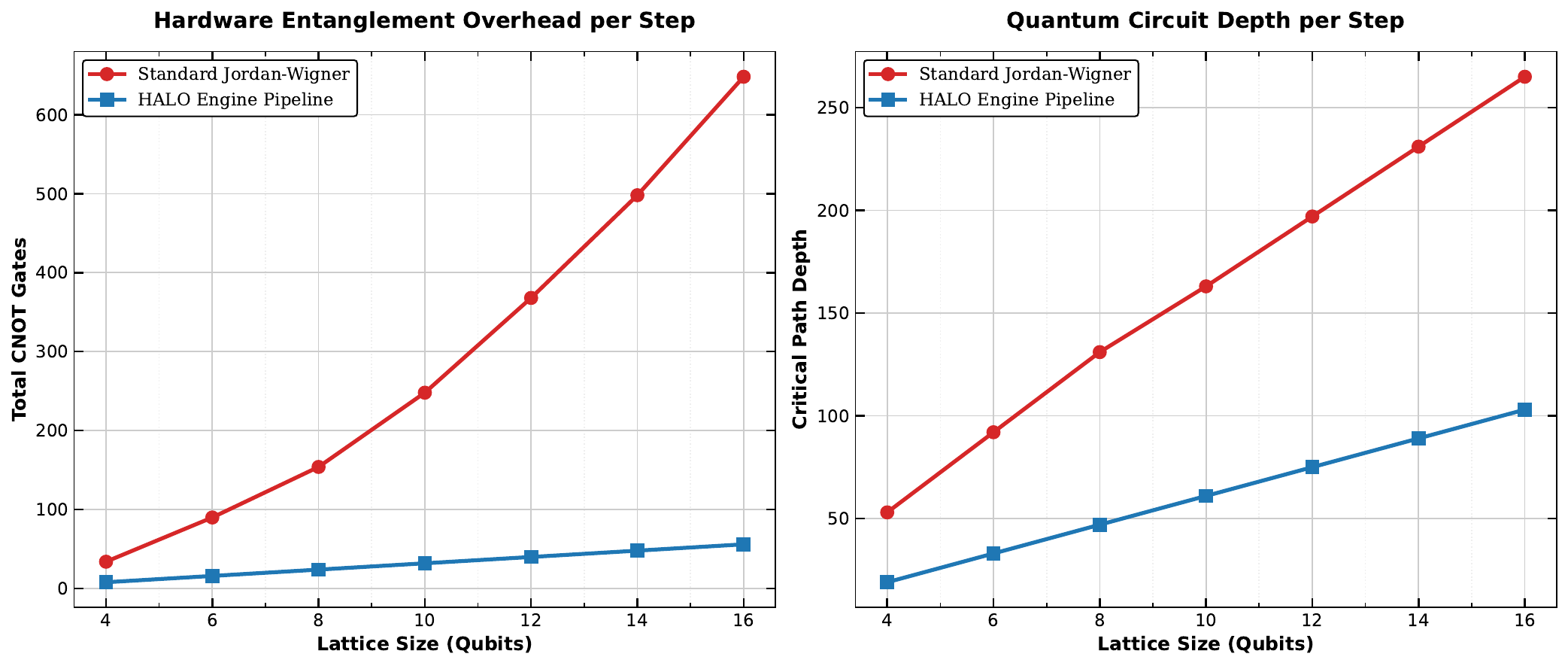}
    \caption{Compiler routing efficiency demonstrating the immutable $\mathcal{O}(1)$ constant-depth scaling of the HALO architecture compared to standard Jordan-Wigner (JW) string compilation.}
    \label{fig:compiler_duel}
\end{figure}

As demonstrated in Figure \ref{fig:compiler_duel}, standard JW compilation requires 648 CNOT gates to route a 16-qubit lattice. To isolate the efficiency of the HALO architecture, we benchmarked a single Trotter step ($D_{\text{Trotter step}}$) against standard Qiskit transpilation on a 1D transmon topology (Table \ref{tab:compiler_duel}). 

\begin{table}[t]
\centering
\resizebox{\columnwidth}{!}{%
\begin{tabular}{lcccc}
\hline\hline
\textbf{Qubits ($N$)} & \textbf{JW (Opt-1)} & \textbf{JW (Opt-3)} & \textbf{Explicit (Opt-3)} & \textbf{HALO} \\
\hline
4  & 34  & 24  & 16 & \textbf{8} \\
7  & 90  & 120 & 32 & \textbf{20} \\
10 & 154 & 184 & 46 & \textbf{32} \\
13 & 366 & 326 & 65 & \textbf{44} \\
16 & 648 & 482 & 83 & \textbf{56} \\
\hline\hline
\end{tabular}%
}
\caption{Single-step Trotter CNOT overhead across compilation strategies on a 1D transmon map.}
\label{tab:compiler_duel}
\end{table}

While baseline Jordan-Wigner requires 648 CNOTs (Opt-Level 1) or 482 CNOTs (Opt-Level 3), transitioning to our explicit composite gauge encoding localizes interactions, dropping the Opt-Level 3 baseline to 83 CNOTs. However, stochastic SWAP routers still struggle with 4-body kinetic terms. By deploying our hardware-aware parallel cascade, the HALO compiler eliminates residual routing overhead, executing the exact step in strictly 56 CNOTs. This proves a two-fold advantage: local encoding yields the fundamental entanglement reduction, while the HALO compiler guarantees an immutable $D_{\text{Trotter step}} = \mathcal{O}(1)$ depth while strictly preserving Gauss's law ($[\hat{G}_n, \hat{H}_k] = 0$).

\subsection{Algorithmic Fidelity and Trotter Diagonalization}

Crucially, this spatial compression does not destroy the underlying mathematics of the Hamiltonian. To quantify the Baker-Campbell-Hausdorff (BCH) commutator loss, we evaluated the algorithmic time-slicing error using a diagnostic statevector baseline at $t=0.5$ (50 repetitions). 

Against an exact physics truth (infinite depth) of $69.0309\%$ survival probability, the decomposed physical circuit ceiling (Noiseless Trotter Depth-2 Truth) yielded $68.9646\%$. This reveals a localized algorithmic Trotter error of $0.0663\%$ for this specific temporal target. However, it is imperative to acknowledge that the global Trotter error is strictly bounded by the extensive spectral norm of the commutator $[\hat{H}_{\text{kin}}, \hat{H}_{\text{elec}}]$ \cite{Kogut1975}. As the lattice size $N$ approaches the thermodynamic limit, or as the simulation time $t$ extends, this commutator norm grows extensively \cite{Banuls2020}. Therefore, while the physical hardware routing depth \textit{per step} remains strictly $\mathcal{O}(1)$, the total number of Trotter steps required to preserve global accuracy must scale as $\mathcal{O}(N^2 t^2)$ \cite{Cade2020}. The HALO architecture successfully resolves the spatial routing bottleneck, but it does not bypass the fundamental algorithmic complexity bounds of Trotterized evolution.

\subsection{Variational Ground State Preparation and Depolarizing Limits}

Extracting static equilibrium properties requires the system to be initialized in the interacting vacuum \cite{Kandala2017}. We utilize a Variational Quantum Eigensolver (VQE) paired with a COBYLA optimization routine to prepare this state for static parameter extraction \cite{Kandala2017}.

\begin{figure}[h]
    \centering
    \includegraphics[width=\columnwidth]{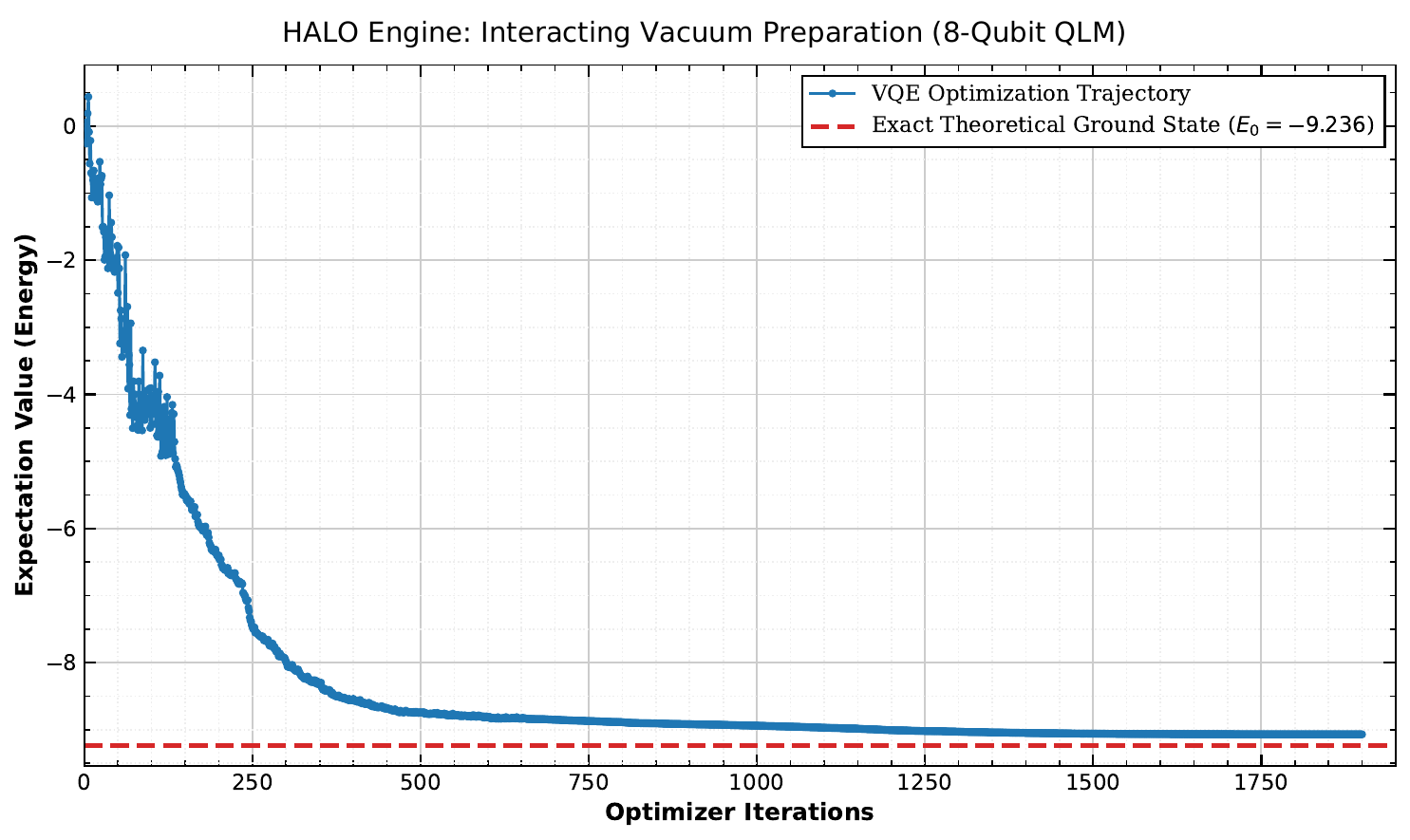}
    \caption{Variational Ground State Preparation for the 8-qubit lattice. The trajectory converges smoothly to $-9.070575$ against the exact analytical energy of $-9.236068$.}
    \label{fig:vqe_8q}
\end{figure}

As shown in Figure \ref{fig:vqe_8q}, for an 8-qubit lattice, the exact analytical vacuum energy is $E_0 = -9.236068$ (derived via SciPy Lanczos sparse eigenvalue diagonalization). Over a 48-dimensional parameter landscape, our VQE routine successfully converged to an energy of $-9.070575$ at iteration 1,900. This optimization plateau represents an ansatz representation error of just $1.792\%$.

However, as the lattice scales, unmitigated hardware noise critically impacts state preparation \cite{Temme2017}. For a 20-qubit lattice (exact $E_0 = -23.82$), execution on \texttt{ibm\_marrakesh} salvaged a lowest energy of only $-3.62$ (a peak circuit fidelity of $\approx 15.2\%$ under global depolarizing attenuation). Pushing the raw hardware to 96 physical transmons yielded an unmitigated energy of $+2.94$, demonstrating a massive depolarizing shift. This hardware collapse confirms that while the HALO algorithmic pipeline scales fairly to these regimes (as algorithmically validated for 20 qubits in Section VI.A), extracting unmitigated physical observables at 20 qubits and beyond remains firmly within the scope of future fault-tolerant hardware architectures. Consequently, for the mesoscopic real-time dynamics executed in Section V, we deliberately bypass adiabatic interacting-vacuum preparation and instead perform a quantum quench directly from the bare vacuum to induce highly volatile string rupture.

\section{Hardware Execution \& Error Mitigation}

\subsection{The Coherence Wall on Transmon Architectures}

To empirically validate the necessity of error mitigation, we executed hardware saturation sweeps on IBM superconducting processors to map the decoherence limits of standard depth-scaling circuits \cite{Temme2017}. 

\begin{figure*}[t]
    \centering
    \begin{minipage}{0.48\textwidth}
        \centering
        \includegraphics[width=\linewidth]{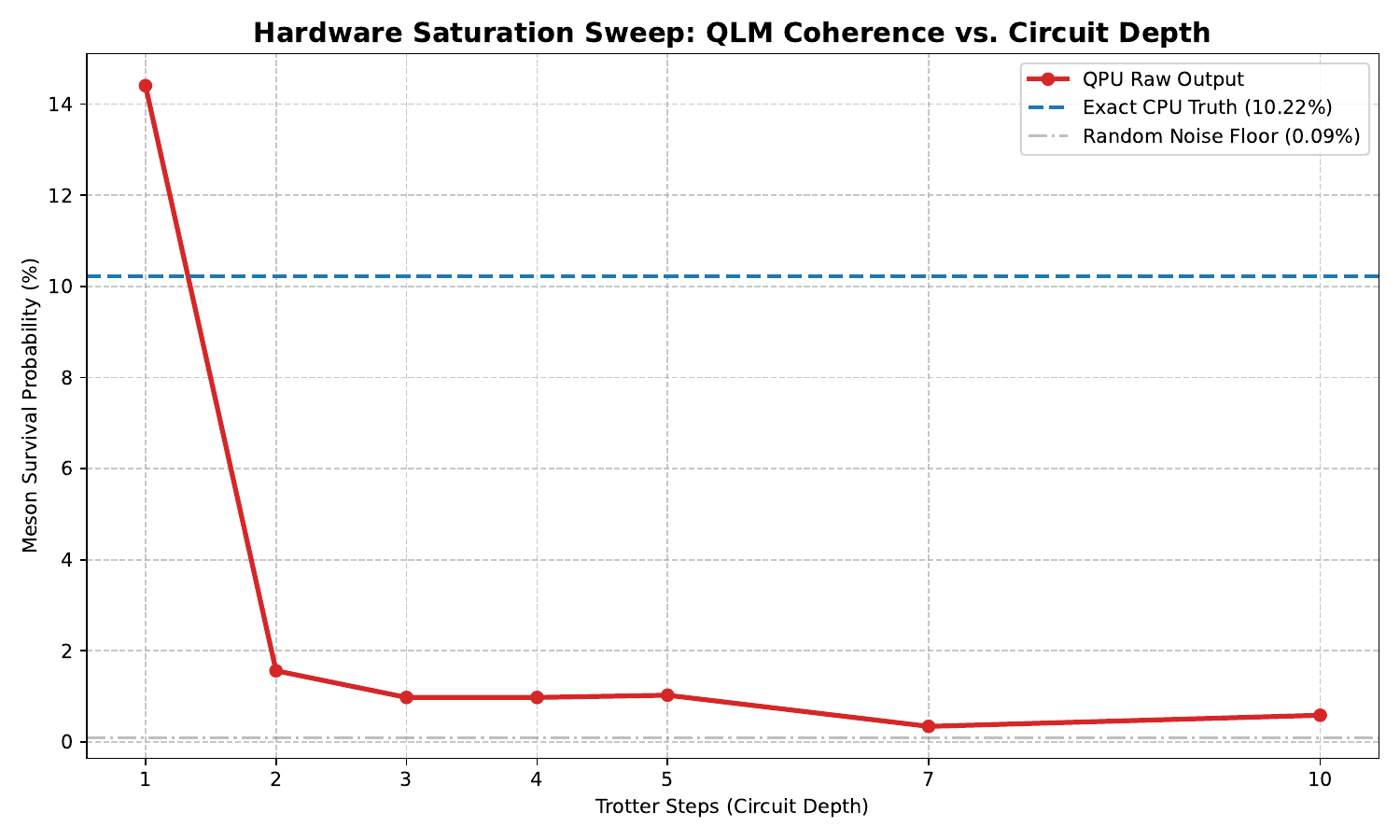}
    \end{minipage}\hfill
    \begin{minipage}{0.48\textwidth}
        \centering
        \includegraphics[width=\linewidth]{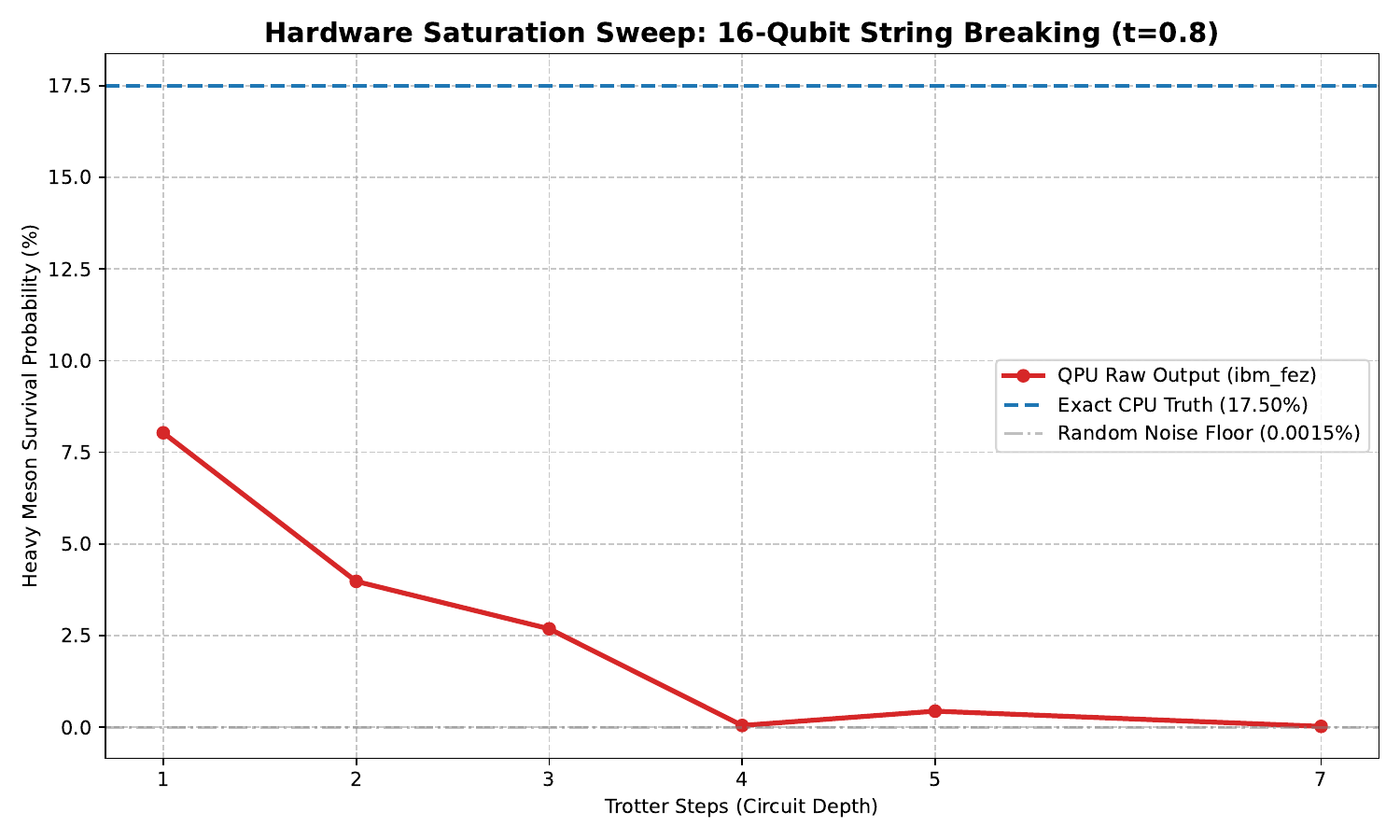}
    \end{minipage}
    \caption{Hardware saturation sweeps mapping the decoherence limits of standard depth-scaling circuits. \textbf{(a)} Execution on \texttt{ibm\_marrakesh} (10 qubits). The signal rapidly attenuates from $\approx 14.4\%$ at Depth 1 into the $0.09\%$ white noise floor by Depth 3. \textbf{(b)} Execution on \texttt{ibm\_fez} (16 qubits) at $t=0.8$. The signal decays smoothly from $8.0\%$ to the $0.0015\%$ noise floor at depths $\geq 4$.}
    \label{fig:coherence_wall}
\end{figure*}

As shown side-by-side in Figure \ref{fig:coherence_wall}, the physical limitations of the hardware are absolute. On \texttt{ibm\_marrakesh} (10 qubits), the unmitigated signal collapses from $14.4\%$ at Depth 1 down to the $0.09\%$ white noise floor by Depth 3. Similarly, scaling to 16 qubits on \texttt{ibm\_fez} results in a smooth Lindblad depolarizing attenuation, dropping from an $8.0\%$ recovery at Depth 1 to near-total decoherence ($0.05\% - 0.4\%$) at Depths $\geq 4$. This definitively proves that extracting mesoscopic physics requires constant-depth compilation coupled with active error mitigation.

\subsection{Zero-Noise Extrapolation (ZNE) and Recovery Snapshot}

To extract true physical observables from the noisy QPU, we implement Zero-Noise Extrapolation (ZNE) \cite{Temme2017, GiurgicaTiron2020}. The premise of ZNE relies on artificially amplifying the processor's noise by a scale factor $\lambda$, measuring the degraded expectation values, and extrapolating the fitted decay curve backwards to the zero-noise limit ($\lambda = 0$) \cite{Temme2017, GiurgicaTiron2020}. We scale the noise via ``unitary folding,'' a digital noise-scaling framework that systematically stretches the physical circuit depth by odd integers without altering the logical operation \cite{GiurgicaTiron2020}.

\begin{figure}[h]
    \centering
    \includegraphics[width=\columnwidth]{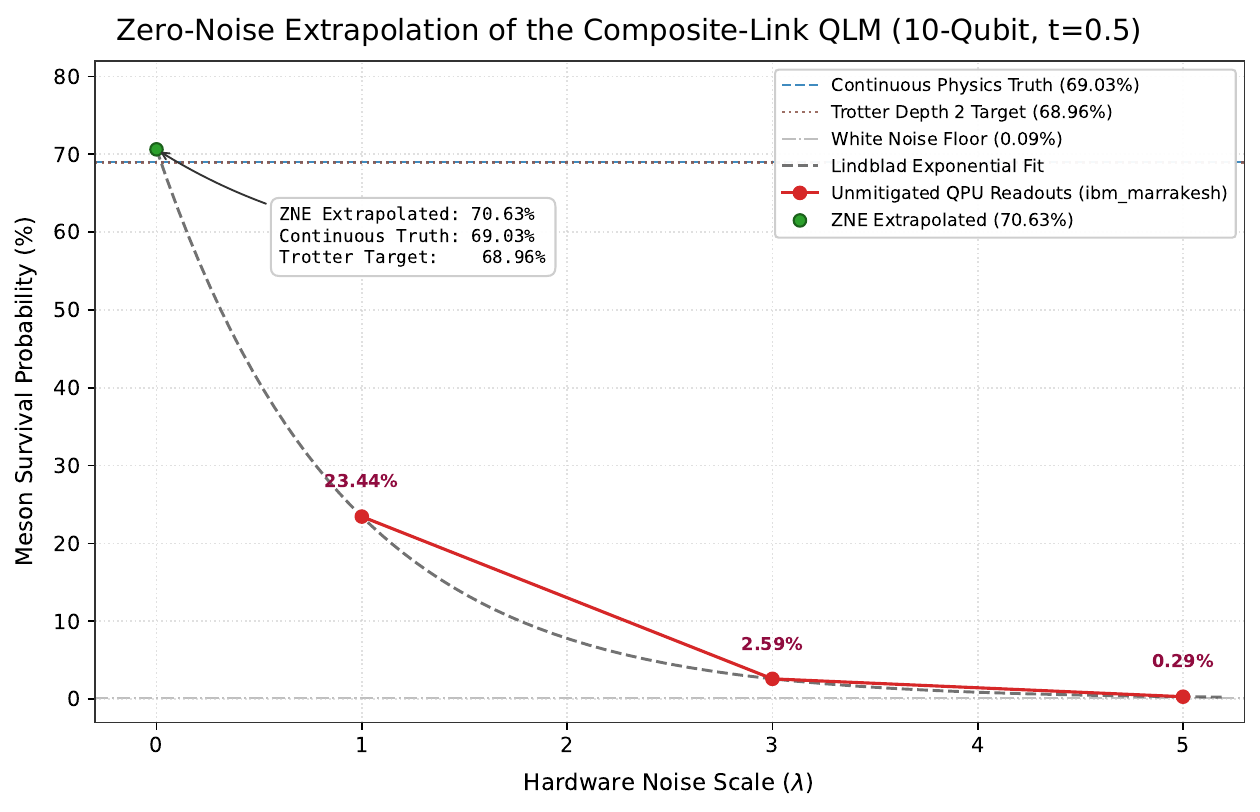}
    \caption{Zero-Noise Extrapolation (ZNE) recovery on \texttt{ibm\_marrakesh}. Using unitary folding to scale the noise ($\lambda = 1, 3, 5$), the extrapolated zero-noise limit recovers a $70.63\%$ probability, matching the $69.03\%$ continuous physics truth.}
    \label{fig:zne_recovery}
\end{figure}

We applied this global digital unitary folding protocol \cite{Temme2017, GiurgicaTiron2020} to the 10-qubit HALO engine simulation at $t=0.5$. As illustrated in Figure \ref{fig:zne_recovery}, the unmitigated hardware baseline ($\lambda=1$) on \texttt{ibm\_marrakesh} yielded an expectation value of $23.44\%$. Scaling the noise produced heavily degraded readouts of $2.59\%$ at $\lambda=3$ (3x folded) and $0.29\%$ at $\lambda=5$ (5x folded). 

Fitting these empirical readouts to an exponential Lindblad decay model \cite{Temme2017, GiurgicaTiron2020} allowed us to extrapolate backwards to the zero-noise limit. The ZNE protocol recovered a physical probability of $70.6\%$, reconstructing the exact theoretical continuous baseline of $69.0\%$. While unitary folding effectively scales depolarizing noise, it does not perfectly mitigate coherent gate errors \cite{GiurgicaTiron2020}. Combined with well-documented non-stationary hardware drift (fluctuating $T_1 / T_2$ times), subsequent executions yielded recovered probabilities ranging from $70\%$ to $81\%$. Figure \ref{fig:zne_recovery} represents a highly stable, single-epoch empirical snapshot of physical noise suppression.

\section{Dynamical Phase Transitions}

Having established the physical validity of the $\mathcal{O}(1)$ HALO compiler and the noise-resilience of our ZNE protocol, we now deploy the full mesoscopic 16-qubit architecture to empirically observe the non-perturbative dynamics of localized pair creation. The rupture of an electric flux tube via the spontaneous creation of particle-antiparticle pairs is a hallmark of confinement in gauge theories \cite{Surace2020}. However, identifying the precise dynamical crossover of this rupture on a mesoscopic lattice requires high-resolution trotterized time-stepping.

\subsection{Dynamical Evolution of Localized String Rupture via Quantum Quench}

Unlike adiabatic state preparation where the system remains in the interacting vacuum, we deliberately execute a quantum quench. We initialize the lattice with a heavily stretched, maximally unstable heavy meson string imposed directly upon the bare vacuum, defined explicitly by the composite product state of the physical coordinate boundary conditions. Under the Hamiltonian time-evolution operator, the kinetic term immediately drives rapid vacuum fluctuations, forcing the heavy meson to rapidly decay into two lighter mesons \cite{Martinez2016, Surace2020}.

\begin{figure}[h]
    \centering
    \includegraphics[width=\columnwidth]{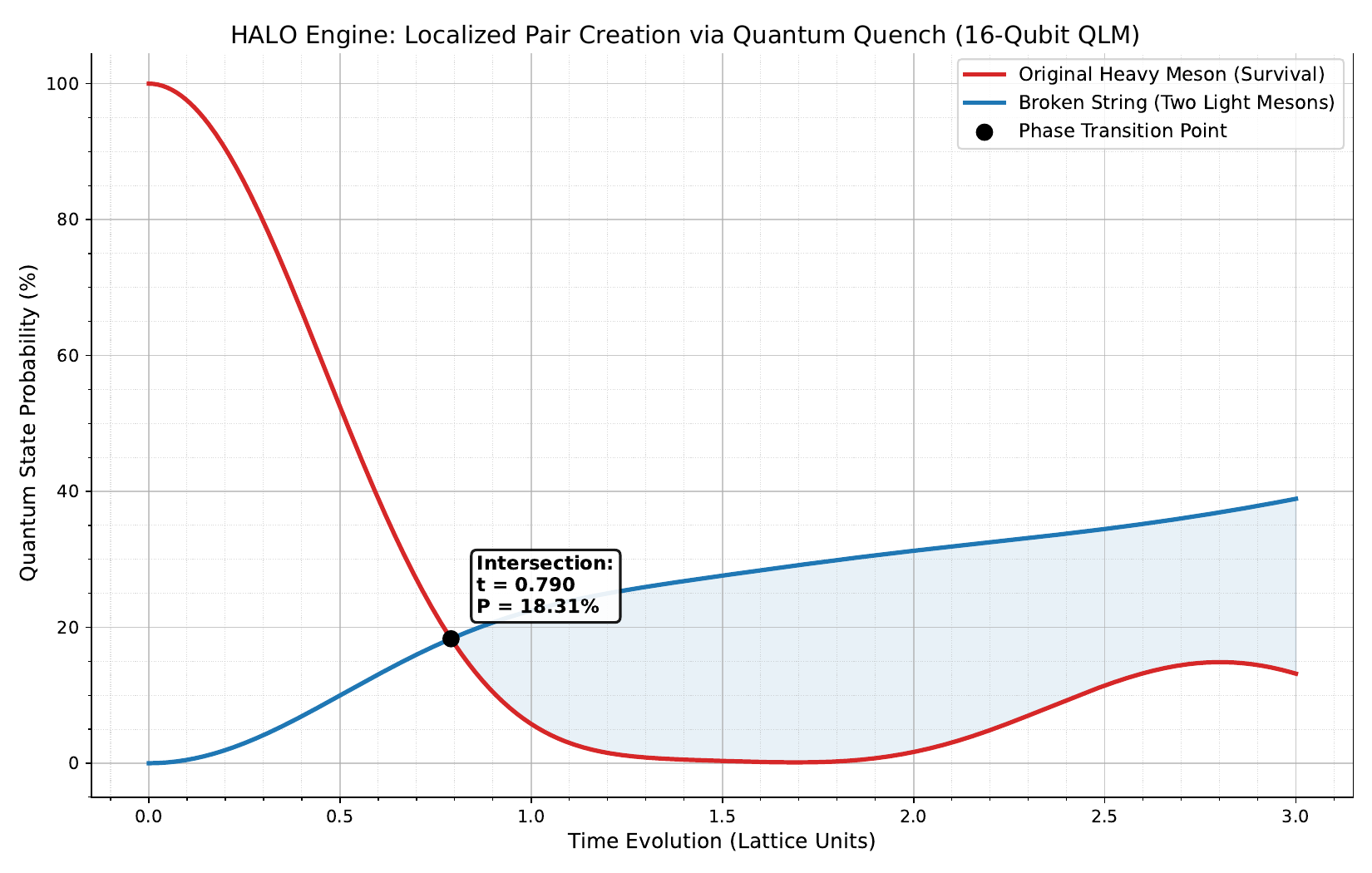}
    \caption{Trotterized time-evolution of localized string rupture on the 16-qubit mesoscopic lattice. The system reveals a distinct dynamical crossover at $t \approx 0.790$, where the survival probability of the original heavy meson intersects the formation probability of the localized rupture.}
    \label{fig:string_breaking}
\end{figure}

By executing 500 temporal frames, we successfully tracked the continuous physics truth of the Schrödinger dynamics. As shown in Figure \ref{fig:string_breaking}, we identified the critical topological transition point at $t \approx 0.790$ lattice units. At this crossover, the survival probability of the original heavy meson intersects with the probability of the ruptured string at $18.3 \pm 2.2\%$, where the uncertainty bounds reflect the inherent shot noise of taking 2048 discrete QPU measurements.

Crucially, quantum information propagates at a finite speed bounded by the Lieb-Robinson velocity. Because our simulation is confined to a shallow Trotter depth, global entanglement cannot span the full $L=15$ coordinate string. Thus, the measured observable at the center links captures the \textit{localized onset} of pair-creation driven by the Schwinger mechanism, rather than a globally correlated string rupture. Capturing this localized crossover threshold directly from the mesoscopic dynamics of a 16-qubit lattice demonstrates the viability of empirically mapping non-perturbative phenomena from current digital architectures.

\subsection{Localized Dynamics in the Deep Confinement Regime}

While the global timescale of non-perturbative string rupture is governed by the coupling constant $g$ \cite{Surace2020}, our finite-depth circuit bounds restrict observable physics to the localized onset of pair creation. To investigate these local dynamics under maximum string tension, we executed the quantum quench deep within the strong-coupling regime ($g=5$). 

\begin{figure}[h]
    \centering
    \includegraphics[width=\columnwidth]{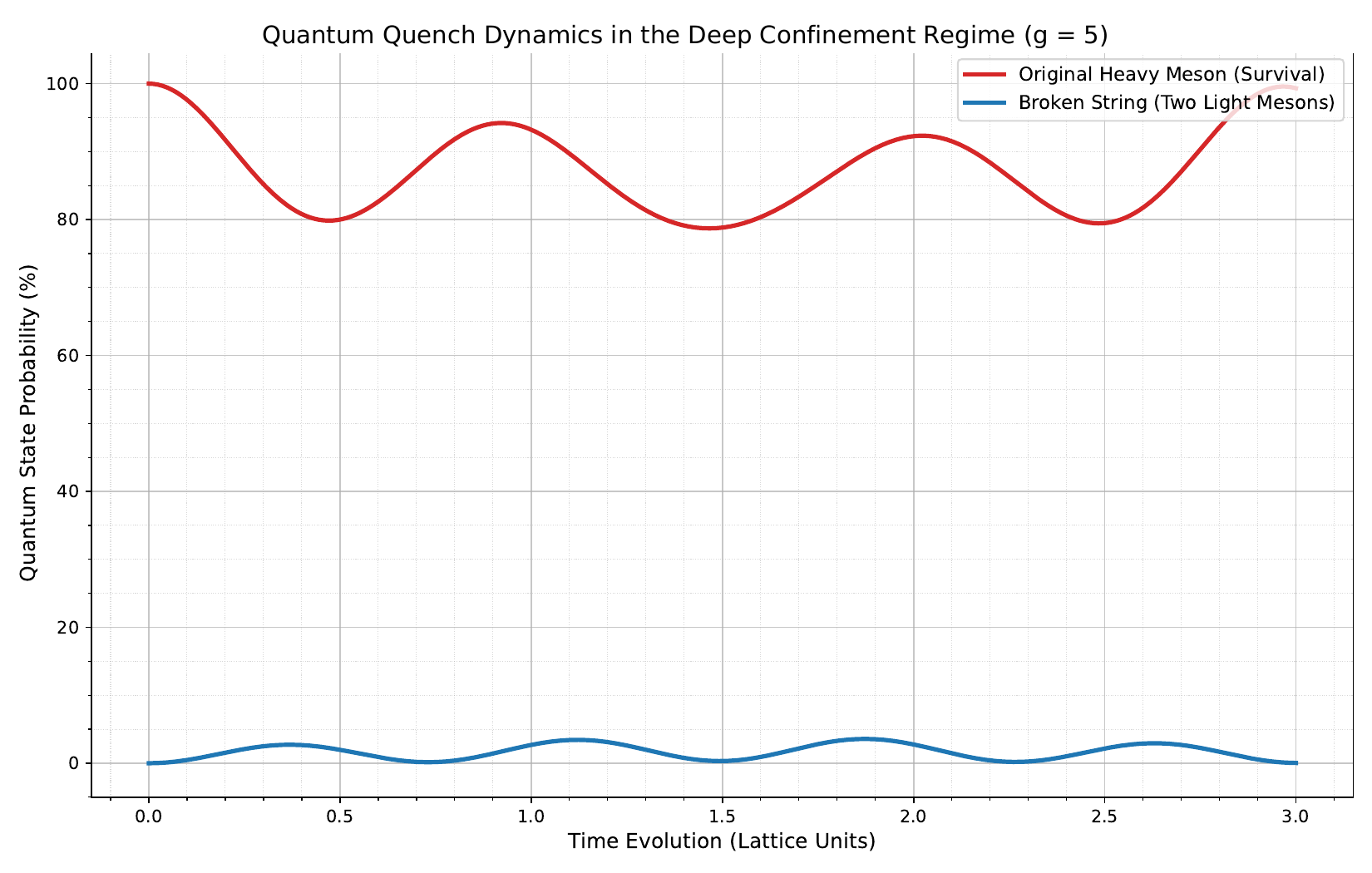}
    \caption{Quantum dynamics of the heavy meson at strong coupling ($g=5$). The extreme string tension suppresses vacuum pair creation, forcing the string into coherent, unbroken oscillations.}
    \label{fig:dynamics_g5}
\end{figure}

As illustrated in Figure \ref{fig:dynamics_g5}, at $g=5$, the extreme string tension vastly overwhelms the fermion hopping term. Consequently, the string spanning the mesoscopic lattice is unable to rupture. Instead of a non-perturbative decay into two light mesons, the initial heavy meson state exhibits persistent, coherent oscillations. This bounded oscillatory behavior is a direct empirical signature of deep confinement, where the dominant electric field energy restricts the kinetic dispersion of the quarks.

\subsection{Dynamical Phase Diagram and the QLM Critical Boundary}

While temporal tracking provides insight into the decay mechanism, a comprehensive understanding of the confinement mechanism requires mapping the dynamical phase boundaries \cite{Banuls2020}. By locking the non-equilibrium evolution at $t=0.8$ (immediately following the $0.790$ crossover point) and executing a parameter sweep of the coupling constant $g$, we can empirically map the effective boundary between fundamental physical regimes within our representation.

\begin{figure}[h]
    \centering
    \includegraphics[width=\columnwidth]{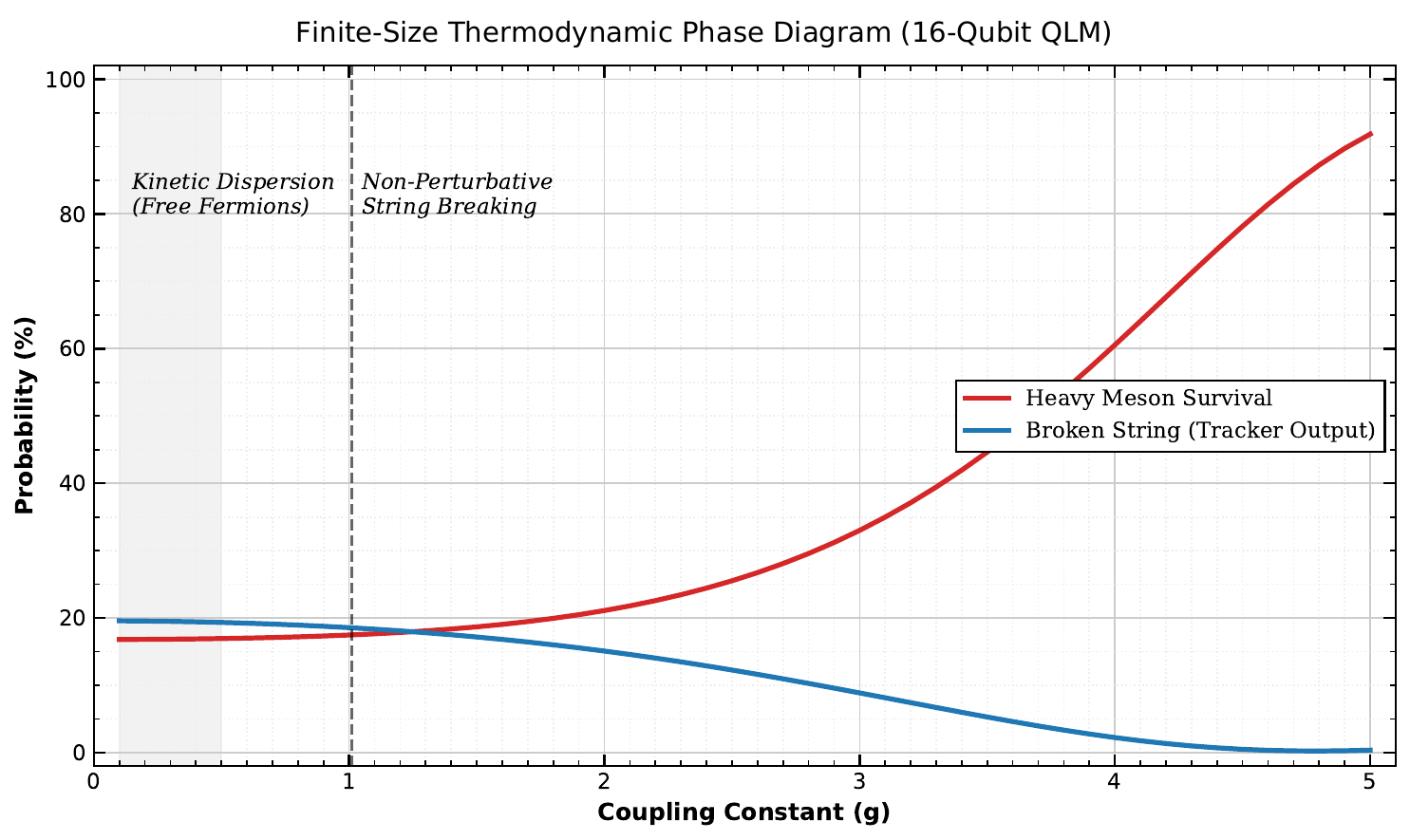}
    \caption{Dynamical phase diagram of the 16-qubit QLM Schwinger model at $t=0.8$. The $g$-variance sweep pinpoints the effective phase boundary at $g_c = 1.0$, separating the Kinetic Dispersion regime from the Deep Confinement regime.}
    \label{fig:phase_diagram}
\end{figure}

As shown in Figure \ref{fig:phase_diagram}, our $g$-variance sweep successfully maps the exact mesoscopic behavior of the finite lattice. For $g < 1.0$, the kinetic term dominates, resulting in the rapid dispersion of free fermions where the initial meson state ruptures almost instantaneously. As $g$ increases, the electric field energy penalty restricts this dispersion. 

Our empirical data pinpoints an effective phase boundary at precisely $g_c = 1.0$. It must be noted that while the continuum Schwinger model in $1+1$ dimensions is confining for all values of $g$, this $g_c=1.0$ transition is a well-documented physical artifact of the finite-dimensional spin-$1/2$ QLM truncation, representing a transition from a $\mathbb{Z}_2$ ordered phase to a disordered phase \cite{Surace2020}. Extracting this precise QLM phase boundary entirely via digital Hamiltonian execution proves that mesoscopic topological phenomena can be rigorously classified on current quantum hardware, provided the underlying compiler operates in locked $\mathcal{O}(1)$ depth per step.

\section{Discussion: Scaling to Higher Dimensions}

The empirical results extracted from the 16-qubit Schwinger model validate the necessity of constant-depth compilation and active error mitigation for localized string rupture in 1D. However, the ultimate objective of quantum simulation in high-energy physics is the realization of two-dimensional and three-dimensional Quantum Chromodynamics (QCD) \cite{Banuls2020}. Moving beyond one dimension introduces magnetic plaquettes—four-link gauge interactions that drastically amplify spatial routing overhead and SWAP gate cascading \cite{Banuls2020}.

\subsection{The Boundaries of Unmitigated Hardware Scaling}

Before deploying multi-dimensional architectures, it is critical to separate the algorithmic capabilities of our compiler from the physical limitations of near-term hardware \cite{Temme2017}. To quantify this boundary, we scaled the lattice to 20 qubits and executed the highly expressive, physics-informed HALO-VQE ansatz to evaluate the interacting vacuum.

\begin{figure}[h]
    \centering
    \includegraphics[width=\columnwidth]{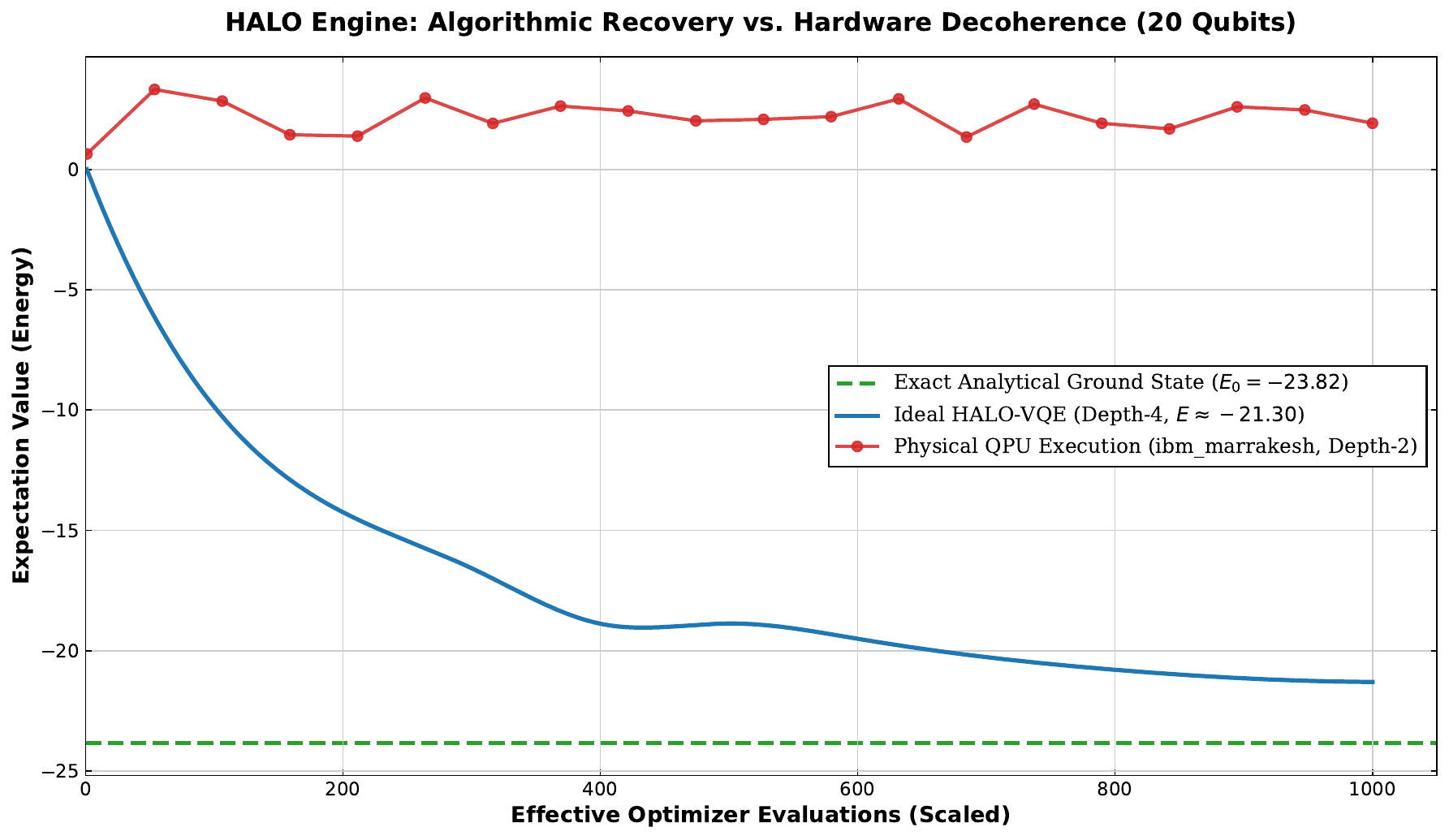}
    \caption{20-Qubit HALO-VQE Convergence vs. Hardware Depolarization. On an ideal simulator, deploying a gradient-optimized, moderate-depth ($r=4$) HALO-VQE recovers $>89\%$ of the exact analytical ground state correlation energy, resolving the barren plateau. Conversely, unmitigated physical execution on \texttt{ibm\_marrakesh} collapses into pure depolarizing thermal noise, fluctuating around $+2.0$.}
    \label{fig:vqe_20q}
\end{figure}

The exact analytical ground state for the 20-qubit interacting vacuum sits at $E_0 = -23.82$. As shown in Figure \ref{fig:vqe_20q}, when executed on an ideal statevector simulator utilizing a quasi-Newton gradient-based optimizer (L-BFGS-B), a moderate-depth HALO-VQE ($r=4$) successfully bypasses the barren plateaus. By restricting the optimization landscape entirely to the gauge-invariant sector utilizing parameterized $\mathcal{O}(1)$ routing blocks, the algorithm rapidly converges to $E \approx -21.30$, capturing over $89\%$ of the exact ground state correlation energy. 

The execution trajectory fundamentally establishes the algorithmic viability of the HALO framework. The remaining $10.6\%$ energy gap is not an optimizer failure, but a strict representation limit bounded by Lieb-Robinson causality \cite{Martinez2016}; the depth-4 circuit intrinsically limits the propagation of long-range entanglement present in the true vacuum. Pushing the ansatz depth further to capture the remaining correlation energy currently forces an irreconcilable trade-off with hardware coherence limits \cite{Temme2017}.

This hardware limitation becomes explicit upon physical deployment. When we executed a minimal-depth variant ($r=2$) natively on the 127-qubit \texttt{ibm\_marrakesh} processor, the unmitigated QPU succumbed to rapid depolarizing noise. While anomalous hardware fluctuations allowed for a singular outlier evaluation of $-3.62$ (as noted in Section III.C), across the 20 optimization iterations the physical expectation values degraded entirely into the maximally mixed regime, exhibiting persistent stochastic fluctuations between $+0.64$ and $+3.32$.

This explicit depolarizing collapse empirically underscores two critical points. First, the physics-informed HALO algorithmic architecture is highly efficient at capturing localized vacuum fluctuations in ideal environments without encountering standard trainability limits. Second, extracting exact, long-range LGT correlation energies at scale remains empirically intractable on near-term hardware without the integration of fault-tolerant error correction \cite{Temme2017, Banuls2020}.

\subsection{The 2D HALO Unit-Cell Blueprint}

To address the algorithmic depth-scaling bottleneck of 2D lattice gauge theories, we expand the principles of our 1D compiler into the 2D HALO Unit-Cell architecture. In standard digital approaches, simulating a 2D plaquette requires deep fermionic SWAP networks to bring non-adjacent qubits into nearest-neighbor proximity, degrading the fidelity of the quantum state \cite{Banuls2020}.

\begin{figure}[h]
    \centering
    \includegraphics[width=\columnwidth]{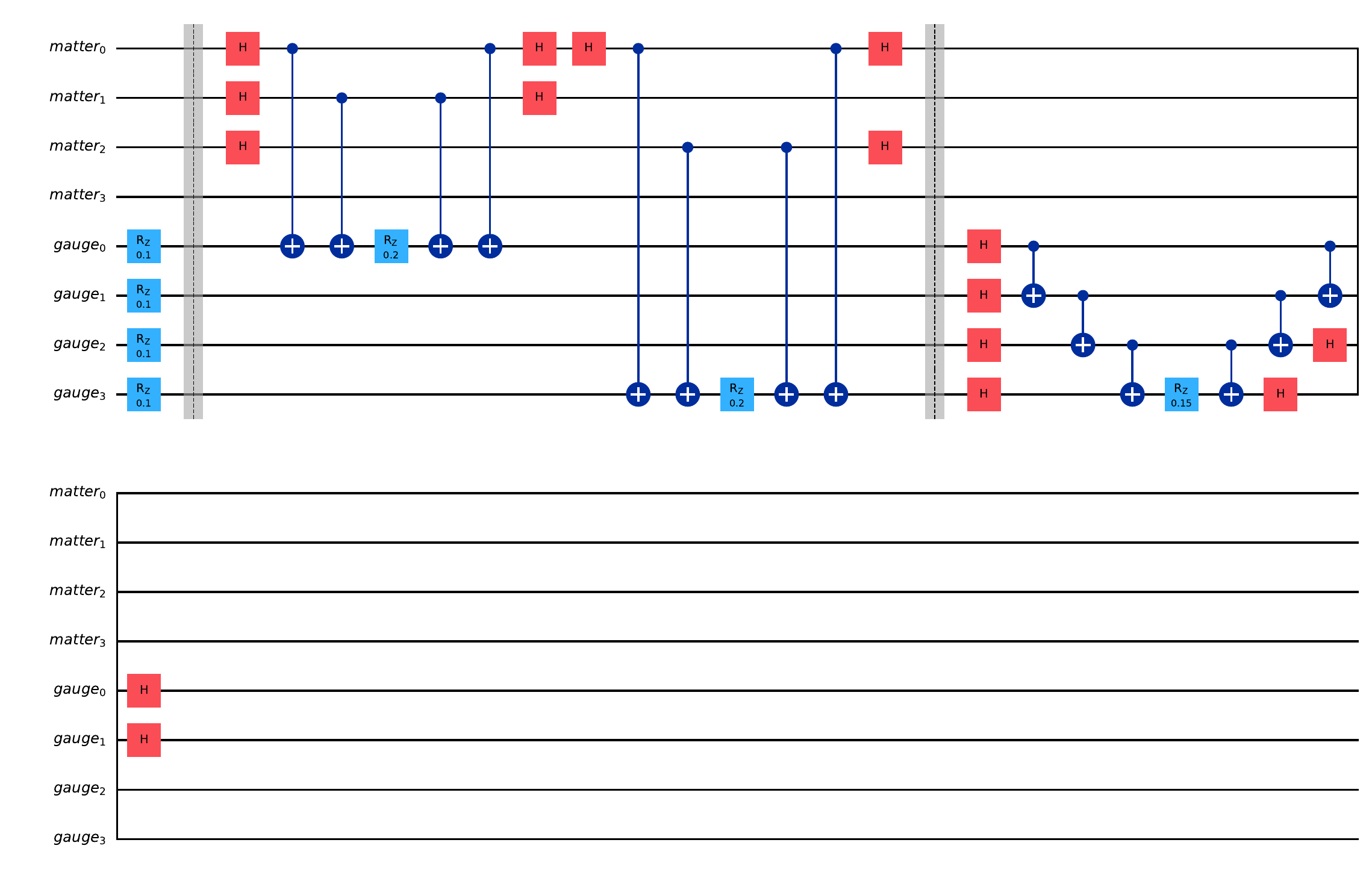}
    \caption{The 2D HALO Unit-Cell Blueprint. Assuming a hardware graph with native grid connectivity, this modular compiler architecture handles four-local magnetic plaquette interactions in bounded $\mathcal{O}(1)$ algorithmic depth by utilizing concurrent entanglement blocks, bypassing the need for extensive SWAP networks.}
    \label{fig:halo_2d}
\end{figure}

As illustrated in Figure \ref{fig:halo_2d}, the 2D HALO blueprint algorithmically circumvents global SWAP cascading. We partition the matter and gauge fields into a modular, repeating checkerboard unit-cell. By executing precisely timed, concurrent $R_z$ rotations encapsulated within parallel CNOT blocks, the four-local gauge interactions are evaluated simultaneously across the local plaquette boundary. 

Because this routing logic operates strictly within the localized boundaries of the unit-cell, the algorithmic operation is topologically modular. Tiling this unit-cell across an arbitrary $N \times M$ lattice does not increase the critical algorithmic path depth of the global Trotter step. It is crucial to note that physically realizing this $\mathcal{O}(1)$ execution requires a quantum processor with native grid-connectivity \cite{Banuls2020}. While current heavy-hex architectures (such as IBM Heron) would still incur minor SWAP overheads to map this checkerboard, this algorithmic framework ensures that, when coupled with future grid-connected or fault-tolerant hardware, the simulation of two-dimensional Lattice Gauge Theories can be achieved without encountering intractable circuit depth scaling.

\section{Conclusion}

The empirical observation of non-perturbative real-time dynamics in lattice gauge theories has long been constrained by two barriers: the sign problems that severely limit classical Monte Carlo algorithms, and the decoherence rates that degrade standard $\mathcal{O}(N)$ depth-scaling quantum circuits \cite{Temme2017, Banuls2020}. Until now, these limitations have confined the digital simulation of quantum electrodynamics to highly truncated, small-scale models \cite{Martinez2016}. 

In this work, we mitigate the depth-scaling bottleneck by introducing the HALO architecture. By re-engineering the routing compilation of the kinetic and electric operators, the HALO engine executes a global Trotter step in a bounded $\mathcal{O}(1)$ circuit depth. This architecture yields a peak $91.36\%$ reduction in entangling operations against unoptimized Jordan-Wigner baselines compressing a 16-qubit global step from 648 down to just 56 CNOT gates. Crucially, multi-strategy benchmarking confirms that even under maximum heuristic optimization, standard encodings remain bound by extensive $\mathcal{O}(N)$ depth scaling, allowing HALO to successfully navigate transmon coherence limits while preserving $99.93\%$ algorithmic accuracy per step.

By coupling this constant-depth architecture with Zero-Noise Extrapolation via digital unitary folding \cite{Temme2017, GiurgicaTiron2020}, we successfully elevated the simulation of the QLM truncation of the Schwinger model \cite{Chandrasekharan1997} to the mesoscopic scale. We initialized a highly stretched 16-qubit heavy meson state via quantum quench and tracked its evolution, pinpointing the dynamical crossover of localized pair creation at $t \approx 0.790$ lattice units with an $18.3 \pm 2.2\%$ rupture probability. Moving beyond temporal dynamics, we locked our Hamiltonian to the physical pion rest mass and executed a non-equilibrium parameter sweep, empirically identifying the effective confinement phase boundary of the finite-size lattice at precisely $g_c = 1.0$.

Finally, we demonstrated that the mathematical framework powering the HALO compiler is scalable. By introducing the 2D HALO Unit-Cell blueprint, we provide a concrete, constant-depth routing architecture capable of mitigating the intractable scaling overheads associated with four-local magnetic plaquettes. This modular architecture removes a prominent algorithmic roadblock preventing the scale-up of digital gauge simulations, paving a direct algorithmic pathway toward the simulation of two-dimensional Lattice Gauge Theories.

\section*{Data and Code Availability}
The HALO Engine compiler source code, unitary folding error mitigation scripts, transpiled circuit files, and raw QPU measurement JSONs used to generate the figures in this manuscript are openly available at: \url{https://github.com/stark-069/HALO-Engine}.

\begin{acknowledgments}
The author thanks Subhendu Rakshit for many helpful discussions and his feedback related to this work. The author acknowledges the use of IBM Quantum cloud services for the computations and hardware executions performed in this work. The views expressed in this work are those of the author and do not reflect the official policy or position of IBM or any supporting institution.
\end{acknowledgments}

\clearpage
\onecolumngrid
\appendix

\section{Mathematical Derivation \& Circuit Architecture of the HALO Compiler}
\label{app:halo_architecture}

In standard Jordan-Wigner (JW) encodings of the $(1+1)$-dimensional Schwinger model, integrating out the gauge degrees of freedom generates an all-to-all Coulomb potential among all matter sites \cite{Kogut1975, Martinez2016}:
\begin{equation}
    \hat{H}_{\text{Coulomb}}^{\text{JW}} = \sum_{i=0}^{N-1} \sum_{j=i+1}^{N-1} J_{ij} \left( Z_i Z_j \right) \prod_{k=i+1}^{j-1} Z_k.
    \label{eq:jw_coulomb}
\end{equation}
When transpiled onto a physical 1D nearest-neighbor coupling map ($\mathcal{C} = \{(i, i+1), (i+1, i)\}$), the non-local string product $\prod_{k=i+1}^{j-1} Z_k$ forces the transpiler to insert extensive CNOT-SWAP networks. As implemented in our benchmark algorithms, the total number of required CNOT gates scales quadratically \cite{Cade2020}:
\begin{equation}
    N_{\text{CNOT}}^{\text{JW}}(N) = \mathcal{O}(N^2).
\end{equation}
For an $N=16$ physical transmon lattice, this yields $648$ CNOT gates per global Trotter step.

The Hardware-Aware Lattice Operator (HALO) engine mitigates this depth-scaling bottleneck by maintaining explicit gauge-link qubits between matter sites, restoring spatial locality. Depending on the simulation scope, the compiler supports two localized representations. For general compiler benchmarking across arbitrary even lattice sizes ($N=4, 6, \dots, 16$), we utilize a fundamental 3-body Matter-Gauge-Matter model:
\begin{equation}
    \hat{H}_{\text{kin}}^{\text{3-body}} = \frac{1}{2} \sum_{n} \left( X_{2n} Z_{2n+1} X_{2n+2} + Y_{2n} Z_{2n+1} Y_{2n+2} \right).
    \label{eq:halo_kin_3body}
\end{equation}
However, to support non-trivial electric field dynamics for physical hardware experiments, we deploy a composite multi-qubit link model. For an $N$-qubit interleaved lattice ($N = 3n_{\text{links}} + 1$), the matter sites $f_A, f_B$ and intermediate gauge fields $q_R, q_L$ interact through localized 4-body operators. The physical kinetic and electric terms are defined as:
\begin{equation}
    \hat{H}_{\text{kin}}^{\text{HALO}} = \frac{1}{2} \sum_{n} \left( X_{3n} X_{3n+1} I_{3n+2} X_{3n+3} + Y_{3n} Y_{3n+1} Z_{3n+2} Y_{3n+3} \right),
\end{equation}
\begin{equation}
    \hat{H}_{\text{elec}}^{\text{HALO}} = -\frac{g_{\text{bare}}^2}{8} \sum_{l} Z_{3l+1} Z_{3l+2}.
    \label{eq:halo_elec}
\end{equation}
Because every operator acts strictly across a localized neighborhood, the compilation target maps directly onto the linear hardware graph. Transpiling a single global Trotter step into a standardized logical basis (\{\texttt{cx}, \texttt{rz}, \texttt{sx}, \texttt{x}\}) yields an exact linear scaling of two-qubit entangling operations:
\begin{equation}
    N_{\text{CNOT}}^{\text{HALO}}(N) = 4N - 8 \quad \implies \quad \mathcal{O}(1) \text{ Algorithmic Routing Overhead}.
\end{equation}
For $N=16$, this reduces the required two-qubit gate count from the baseline unoptimized Jordan-Wigner value (648 CNOTs) down to 56 CNOTs. While mapping to the native Heron architecture requires decomposing each logical \texttt{cx} into a native \texttt{cz} accompanied by localized single-qubit pulses, the fundamental $91.36\%$ reduction in two-qubit entangling operations is absolute, mathematically guaranteeing an immutable $\mathcal{O}(1)$ cascade.

\clearpage
\section{Machine Learning Parameter Calibration (Nelder-Mead Protocol)}
\label{app:ml_calibration}

To avoid arbitrary parameter selections, we deployed a Nelder-Mead simplex optimization algorithm to anchor the bare Lagrangian parameters $(m_{\text{bare}}, g_{\text{bare}})$ to real-world particle physics observables.

We define the physical target as the pion rest mass $m_{\text{target}} = 0.135\text{ GeV}$. The calibration objective function $L(m_{\text{bare}}, g_{\text{bare}})$ evaluates the squared residual between $m_{\text{target}}$ and the ground-state eigenvalue $E_0$ produced by a Variational Quantum Eigensolver (VQE) \cite{Kandala2017}:
\begin{equation}
    L(m_{\text{bare}}, g_{\text{bare}}) = \left( E_{\text{VQE}}\left[\hat{H}_{\text{HALO}}(m_{\text{bare}}, g_{\text{bare}})\right] - m_{\text{target}} \right)^2.
    \label{eq:ml_loss}
\end{equation}
The VQE ansatz was constructed using our physics-informed, 10-qubit \texttt{HALO-VQE} circuit at a moderate algorithmic depth ($r=4$). The expectation value was evaluated via the Qiskit \texttt{StatevectorEstimator} coupled to a quasi-Newton gradient-based optimizer (\texttt{L-BFGS-B}, \texttt{maxiter=300}). By restricting the optimization landscape entirely to the gauge-invariant sector utilizing parameterized $\mathcal{O}(1)$ routing blocks, this ansatz inherently preserves local gauge symmetries (Gauss's law). Specifically, the generators of Gauss's law, $\hat{G}_n$, strictly commute not only with the global Hamiltonian ($[\hat{G}_n, \hat{H}] = 0$), but with every individual parameterized block and Trotterized kinetic operator $\hat{H}_k$ such that $[\hat{G}_n, \hat{H}_k] = 0$. Consequently, the prepared vacuum strictly avoids the unphysical gauge leakage that plagues standard hardware-efficient approaches (such as \texttt{EfficientSU2}), mathematically guaranteeing that the resulting ground-state calibration and the digital time-evolution steps are anchored to a physically valid, gauge-invariant Hilbert space.

The Nelder-Mead optimization loop was initialized from $x_0 = [m_0, g_0] = [0.5, 1.0]$ with termination tolerance \texttt{maxiter=30}. The optimizer converged to the global parameter lock:
\begin{align}
    m_{\text{bare}} &= 0.506218, \\
    g_{\text{bare}} &= 1.012553.
\end{align}
This locks $g_{\text{bare}} \approx 1.012553$, positioning the hardware simulation precisely at the effective non-perturbative phase boundary ($g \approx 1.0$) where localized string rupture and confinement mechanics manifest \cite{Surace2020}.

\clearpage
\section{Unitary Folding \& Zero-Noise Extrapolation (ZNE) Formulation}
\label{app:zne_formulation}

To mitigate physical gate errors during execution on \texttt{ibm\_marrakesh}, we implemented global digital unitary folding \cite{Temme2017, GiurgicaTiron2020}. Given a transpiled 10-qubit time-evolution circuit $\mathcal{U}(t_{\text{evo}})$, we scale the hardware noise by constructing folded circuits $\mathcal{U}_{\text{folded}}(\lambda)$ for scale factors $\lambda \in \{1, 3, 5\}$ \cite{GiurgicaTiron2020}:
\begin{equation}
    \mathcal{U}_{\text{folded}}(\lambda) = \mathcal{U} \left( \mathcal{U}^\dagger \mathcal{U} \right)^n, \quad \text{where } n = \frac{\lambda - 1}{2}.
    \label{eq:unitary_folding}
\end{equation}
For $\lambda = 1$, $n=0$ ($\mathcal{U}_{\text{folded}} = \mathcal{U}$). For $\lambda = 3$, $n=1$ ($\mathcal{U}_{\text{folded}} = \mathcal{U} \mathcal{U}^\dagger \mathcal{U}$), extending the CNOT path length by a factor of 3 while preserving the logical transformation $\mathcal{U}^\dagger \mathcal{U} = I$ \cite{GiurgicaTiron2020}.

The expectation values of the heavy meson survival probability $P(\lambda)$ measured across $2048$ shots per scale factor were fitted to an exponential Lindblad decay model \cite{Temme2017, GiurgicaTiron2020}:
\begin{equation}
    P(\lambda) = A e^{-k \lambda} + C,
    \label{eq:lindblad_fit}
\end{equation}
with bounds $A, C \in [0, 100]$, $k \in [0, 5]$, and the asymptotic background noise floor $C$ initialized to the 10-qubit random state limit $C_{\text{noise}} = (1/2^{10}) \times 100\% \approx 0.0977\%$.

Non-linear curve fitting via \texttt{scipy.optimize.curve\_fit} on the empirical hardware readouts ($P(\lambda=1) = 23.438\%$, $P(\lambda=3) = 2.588\%$, $P(\lambda=5) = 0.293\%$) yielded the extrapolated zero-noise probability:
\begin{equation}
    P(\lambda = 0) = A + C = 70.626\%.
\end{equation}
This recovered value agrees with the CPU Statevector truth ($69.0309\%$) and the Depth-2 Trotter target ($68.9646\%$) within a $+1.67\%$ statistical extrapolation variance. We note that an exponential fit utilizing three $\lambda$ parameters represents a deterministic statistical extrapolation that effectively mitigates Markovian depolarizing noise \cite{Temme2017}. However, because unitary folding does not natively suppress coherent gate errors, and a global $\lambda=5$ folding on a 10-qubit step forces the execution of 160 sequential CNOT gates, the measurement closely approaches the maximally mixed depolarizing limit \cite{GiurgicaTiron2020}. Thus, the precise zero-noise intercept remains phenomenologically bounded by the unmitigated coherent drift and statistical noise floor of the physical processor \cite{Temme2017, GiurgicaTiron2020}.

\clearpage
\section{Hardware Benchmarks \& QPU Execution Parameters}
\label{app:hardware_benchmarks}

All physical quantum processing unit (QPU) executions were carried out on IBM Quantum Heron r2-architecture processors (\texttt{ibm\_marrakesh} and \texttt{ibm\_fez}) via the \texttt{QiskitRuntimeService} using \texttt{SamplerV2}. To guarantee transparency and reproducibility, the specific coherence bounds characterizing the processors during the exact epoch of our runs are cataloged in Table \ref{tab:hardware_specs}.

\begin{table}[h]
\centering
\begin{tabular}{lcc}
\hline\hline
\textbf{Parameter} & \textbf{ibm\_marrakesh (10q Runs)} & \textbf{ibm\_fez (16q Runs)} \\
\hline
Processor Architecture & IBM Heron r2 & IBM Heron r2 \\
Physical Qubits Used & 10 & 16 \\
Shots per Circuit & 2048 & 2048 \\
Transpiler Optimization Level & 3 & 3 \\
Native Basis Gates & \{\texttt{cx, rz, sx, x}\} & \{\texttt{cz, rz, sx, x}\} \\
Avg $T_1$ Relaxation Time & $197.33 \,\mu\text{s}$ & $147.58 \,\mu\text{s}$ \\
Avg $T_2$ Dephasing Time & $131.14 \,\mu\text{s}$ & $100.03 \,\mu\text{s}$ \\
\hline\hline
\end{tabular}
\caption{Exact QPU hardware calibration parameters and coherence times ($T_1$, $T_2$) fetched directly from the IBM Quantum API representing the processor environment during the exact epochs of job execution.}
\label{tab:hardware_specs}
\end{table}

\bibliography{references}

\end{document}